\documentclass[12pt]{iopart}

\usepackage{iopams}
\usepackage{graphicx}
\usepackage{indentfirst}
\begin{document}

\title[Near-Ionization-Threshold Injection]{Generation of Low Absolute Energy Spread Electron Beams in Laser Wakefield Acceleration Using Tightly Focused Laser through Near-Ionization-Threshold Injection}

\author{F Li$^1$, C J Zhang$^1$, Y Wan$^1$, Y P Wu$^1$, J F Hua$^1$, C H Pai$^1$, W Lu$^1$, W B Mori$^2$, C Joshi$^2$}
\address{$^1$Department of Engineering Physics, Tsinghua University, Beijing 100084, China}
\address{$^2$University of California Los Angeles, Los Angeles, CA 90095, United State}

\ead{weilu@tsinghua.edu.cn}

\vspace{10pt}
\begin{indented}
\item[]September 2015
\end{indented}

\begin{abstract}
    An enhanced ionization injection scheme using a tightly focused laser pulse with intensity near the ionization potential to trigger the injection process in a mismatched pre-plasma channel has been proposed and examined via multi-dimensional particle-in-cell simulations. The core idea of the proposed scheme is to lower the energy spread of trapped beams by shortening the injection distance. We have established theory to precisely predict the injection distance, as well as the ionization degree of injection atoms/ions, electron yield and ionized charge. We have found relation between injection distance and laser and plasma parameters, giving a strategy to control injection distance hence optimizing beam's energy spread. In the presented simulation example, we have investigated the whole injection and acceleration in detail and found some unique features of the injection scheme, like multi-bunch injection, unique longitudinal phase-space distribution, \emph{etc.} Ultimate electron beam has a relative energy spread (rms) down to 1.4\% with its peak energy 190 MeV and charge 1.7 pC. The changing trend of beam energy spread indicates that longer acceleration may further lower the energy spread down to less than 1\%, which may have potential in applications related to future coherent light source driven by laser-plasma accelerators.
\end{abstract}

\vspace{2pc}
\noindent{\it Keywords}: laser wakefield acceleration, ionization injection, injection distance, energy spread, plasma channel
%\submitto{\PPCF}
%\maketitle

\section{Introduction}

Impressive progress has been made in recent years in theoretical and experimental studies of laser wakefield accelerators which can boost the trapped electrons to GeV level \cite{leemans2006,kim2013,clayton2010,muggli2004,hogan2005,blumenfeld2007} within centimeter-scale and meter-scale plasma, because of the high acceleration gradient at least three orders of magnitude than that of conventional radio-frequency accelerators. In addition, laser wakefield usually has a spatial scale of tens of microns, which gains advantages for generating microscale electron beams (ultra-short duration and low emittance) over rf accelerators by nature. However, injecting electrons into such compact structure which moves close to speed of light becomes extremely challenging.

In virtue of paramount importance of injection controllability over the ultimate beam qualities of trapped electrons beams, several injection techniques have been demonstrated experimentally or via simulations, including laser-ionization injection \cite{oz2007,pak2010,mcguffey2010,pollock2011,liu2011,hidding2012,bourgeois2013,zeng2015}, wakefield-induced ionization injection \cite{martinez2013}, density transition injection \cite{geddes2008}, colliding pulse injection \cite{esarey1997,li2013,xu2014,chen2014} and external magnetic field injection \cite{vieira2011}. Some of the techniques requiring fs-scale synchronization and $\mu$m-scale alignment among several experimental elements, can remarkably enhance the robustness and controllability of injection process, consequently improving the beam qualities, however at the sacrifice of the experimental simplicity and stability.

Standard ionization injection method, which utilizes a single laser pulse and gaseous mixture target (containing a gas with low ionization potential (IP) and another with higher IP) as the typical experimental setup, is effective and feasible while also providing the controllability of the injection process to a certain extent. However, the conventional parametric configuration only allows a continuous ionization of the high-IP gas species which lasts about a Rayleigh length or even longer. As a result of the non-localized ionization, the output beam energy spread is usually large. In order to fulfill a localized injection, a two-stage setup \cite{pollock2011,liu2011}, containing a gaseous mixture section of several millimeters used for the injection and a longer pure gas section used for the main acceleration, is usually adopted. Two-staged injection experiments usually give beam energy spread no less than 5\%.

To decrease the beam energy spread while retaining the experimental feasibility and controlability, we have ameliorated the conventional ionization injection (we call the improved method near-threshold-ionization (NIT) injection) in this paper and confined the injection to a short distance through adjustment of the laser power, temporal and spatial characters and plasma density.

\section{Physical Picture}

Before presenting the improved injection scheme, we first retrospect the principle of the ionization injection. When an intense, short laser pulse propagates through a gaseous mixture consisting of multiple ionization states, the leading edge of the laser pulse is intense enough to fully ionize the low-IP atoms/ions. The released electrons will form a plasma wakefield under the interaction with laser ponderomotive force. Due to the large difference in ionization potential, the inner-shell electrons of the high-IP atoms/ions will not be delivered until the peak laser intensity reaches. This fraction of electrons are released in the fully formed wake and slip backwards relative to the laser pulse. If they gain enough energy from the acceleration field during slipping to move at the phase velocity of the wake $v_\phi$, they get trapped and the subsequent acceleration begins. The motion of electrons in the wakefield can be described by Hamiltonian mechanics \cite{oz2007}. In the co-moving system of reference with $\xi=z-v_\phi t$ and $v_\phi$ the phase velocity of wakefield, it can be show that the quantity $\mathcal{K}=\gamma m_ec^2-v_\phi p_z-e\Psi(\xi)$ is a constant of motion, where $\gamma$ the Lorentz factor and $p_z$ the axial momentum of the witness electron. Here, we have defined the pseudo-potential $\Psi(\xi)=\Phi(\xi)-v_\phi A_z(\xi)$ with $\Phi$ the scalar potential and $A_z$ the longitudinal vector potential of the wake. Assuming that the initial kinetic energy of electrons immediately after ionization is negligible, the trapping condition can be derived by equating the initial and final values of $\mathcal{K}$
\begin{equation}
    \label{eq:trap_condition}
    \Delta\Psi=\Psi_f-\Psi_i=-\frac{m_ec^2}{e}\left(1-\frac{\gamma_\perp}{\gamma_\phi}\right)
\end{equation}
where $\gamma_\perp=(1+(p/m_ec)^2)^{1/2}$ and $\gamma_\phi=(1-(v_\phi/c)^2)^{-1/2}$. For ultra-relativistic driver ($\gamma_\phi\rightarrow\infty$), Eq. (\ref{eq:trap_condition}) reduces to $\Delta\Psi=-m_ec^2/e$.

%The pseudo-potential $\Psi$ reaches its minimum at the rear of the first bucket and the maximum at where $E_z=0$. So if a wake with $\Psi_{\rm max}>\Psi_{\rm th}=\Psi_{\rm min}+m_ec^2/e$ is excited and a portion of electrons are released adjacent to $\xi(\Psi=\Psi_{\rm max})$, they will easily get trapped. A typical laser wakefield structure and ionization injection scheme are illustrated schematically in Fig. \ref{fig:ion_inj}. The black dotted line denotes the threshold of $\Psi$ for trapping. The electrons born above the trapping criteria, \emph{i.e.} at the region colored in yellow, will get trapped eventually. On the other hand, the region colored in green is where the photo-ionization predominantly occurs. Ionization injection takes place only when these two regions have an intersection.

%\begin{figure}
%    \centering
%    \includegraphics[width=8cm]{fig4.pdf}
%    \caption{\label{fig:ion_inj}A typical laser wakefield structure and ionization injection scheme. The light gray lines shows a laser electric field distribution along its propagation direction and the red line shows the acceleration field distribution of the wake. They are both plotted in normalized unit $m_e\omega_0c/e$. The blue line shows the pseudo-potential $\Psi$ in normalized unit $m_ec^2/e$ with a trapping threshold denoted by the black dotted line. The black line denotes the ionization degree of N$^{5+}$. The region colored in yellow and green are where the trapping criteria is satisfied and photo-ionization predominantly takes place.}
%\end{figure}

\begin{figure}
    \centering
    \includegraphics[width=8cm]{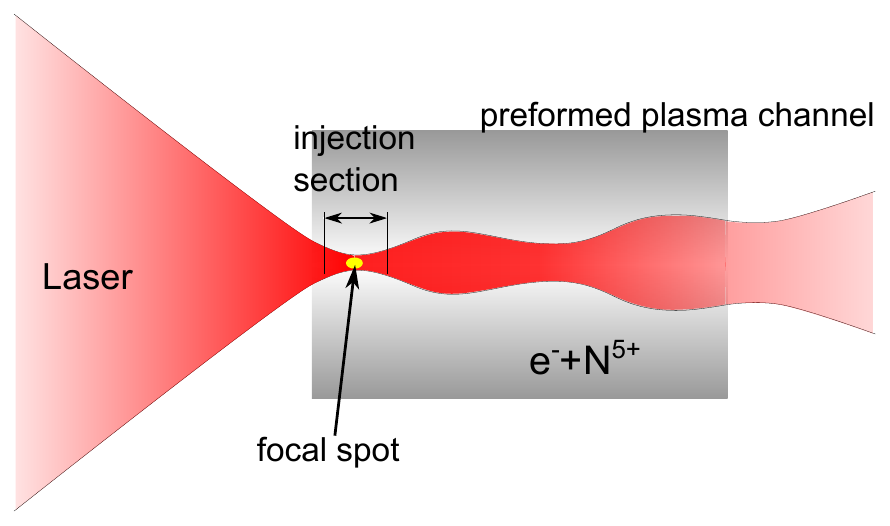}
    \caption{\label{fig:sketch}Schematic of near-threshold ionization injection.}
\end{figure}

Fig. \ref{fig:sketch} illustrates the enhanced ionization injection method. A laser pulse is focused at the upramp edge of a preionized plasma. To shorten the ionization volume, a tightly focused laser spot realized by small f-number focusing optics is essential, for which one can obtain a shorter Rayleigh length $z_R=\pi w_0^2/\lambda_0$ and make the laser diffract rapidly preventing the incessant ionization, where $w_0$ the focal waist and $\lambda_0$ the wavelength of the laser pulse. The wakefield excited at the focal spot is strong enough to trapped the electrons liberated from the high-IP ions, say N$^{5+}$. After the laser diffracting to a larger spot size and the ionization injection stopping, the laser pulse needs to be guided for the subsequent acceleration. Here, we utilized a mismatched plasma channel with a parabolic transverse density profile to limit the laser from over-diffracting and meanwhile allow quick laser defocus near the focal plane.

\section{Absolute Energy Spread and Injection Distance}

The key point of lowering beam's absolute energy spread is to shorten the injection distance $D_{\rm inj}$, making the injection self-truncated. It is intuitively known that the initial absolute energy spread $\Delta\gamma_0 \propto E_{z,{\rm inj}} D_{\rm inj}$, where $E_{z,{\rm inj}}$ is the acceleration field felt by the injected particles which is related to driver strength and injection position. In blowout regime \cite{lu2006}, $E_{z, {\rm inj}}$ can be roughly estimated by \cite{lu2007}
\begin{equation}
    E_{z,{\rm inj}}\sim\frac{m_ec\omega_p}{e}\sqrt{a_0},
\end{equation}
where $\omega_p$ the plasma frequency, $a_0$ the normalized vector potential of laser. Except lowering the initial absolute energy spread, another benefit of shortening $D_{\rm inj}$ is diminishing phase-mixing effect \cite{xu2014} which has been proved the primary factor of beam's emittance degradation during injection.

The laser-ionization process determines $D_{\rm inj}$. For a linearly polarized (LP) or circularly polarized (CP) laser, the ionization rate of the outermost electron from its ground state energy level in the tunneling limit is described by Ammosov-Delone-Krainov (ADK) model \cite{ammosov1986}
\begin{equation}\label{eq:adkmodel}
    W_{\rm ADK} = \frac{4^{n^*}\xi_i}{n^*\Gamma(2n^*)}\left(\frac{2\xi_0}{|E|}\right)^{2n^*-1} \exp{\left(-\frac{2\xi_0}{3|E|}\right)}.
\end{equation}
Here, all the quantities in Eq. (\ref{eq:adkmodel}) (as well as the remaining of the section) are in atomic units. $\xi_i$ is the IP normalized to atomic energy scale $\xi_a\approx27.2\ {\rm eV}$. $E$ is the laser field normalized to atomic field $E_a\approx5.14\times10^{11}\ {\rm V/m}$. $\xi_0=(2\xi_i)^{3/2}$ is a dimensionless characteristic parameter. The effective principal quantum number is defined as $n^*\equiv Z/(2\xi_i)^{1/2}$ where $Z$ is the net charge after ionization. $\Gamma(x)$ is the standard Gamma function.

To find out what determines the injection distance, we start from the electron generation rate as the ionization laser propagates through neutral/ion gas
\begin{equation}\label{eq:electron_rate}
    \partial_t(n/n_g)=[1-(n/n_g)]W_{\rm ADK},
\end{equation}
where $n$ and $n_g$ are generated electron density and neutral/ion gas density respectively. The denotation $n/n_g$ denotes the ionization degree of a single atom/ion. After a laser pulse sweeps across, the ionization degree is
\begin{equation}\label{eq:ion_degree}
    n/n_g=1-\exp(-I(r,z)),
\end{equation}
where the integral $I(r,z)\equiv\int_{-\infty}^{+\infty}W_{\rm ADK}(t)dt$. In weak ionization limit, \emph{i.e.} $I\ll1$, the ionization degree $n/n_g$ approximately equals to $I$. Consider a LP or CP laser pulse with bi-gaussian profile, the electric field is
\begin{eqnarray}\label{eq:laserfield}
    E&=\frac{w_0}{w}E_p\exp\left(-\frac{r^2}{w^2}-\frac{\xi^2}{\sigma_\xi^2}\right) \quad &\mbox{for CP}, \\
    E&=\frac{w_0}{w}E_p\exp\left(-\frac{r^2}{w^2}-\frac{\xi^2}{\sigma_\xi^2}\right)\cos(k\xi) \quad &\mbox{for LP},
\end{eqnarray}
where $E_p,\ w,\ w_0,\ \sigma_\xi$ and $k$ are laser's peak field, transverse size, focal waist, pulse length and wave number respectively. Combining Eq. (\ref{eq:adkmodel}) and (\ref{eq:laserfield}), integrating $W_{\rm ADK}$ (see \ref{appendix:A}) over $t$ yields the on-axis ionization degree in weak ionization limit
\begin{equation}\label{eq:onaxis_iondeg}
    n/n_g\simeq I(0,z)=
    \cases{\frac{\kappa}{c}\sqrt{\pi}\sigma_\xi\left(\frac{\hat{w}}{\epsilon_\psi}\right)^{2n^*-\frac{3}{2}} e^{-\hat{w}/\epsilon_\psi} \quad \mbox{for CP}, \\
        \frac{\kappa}{c}\sqrt{2}\sigma_\xi\left(\frac{\hat{w}}{\epsilon_\psi}\right)^{2n^*-2} e^{-\hat{w}/\epsilon_\psi}, \quad \mbox{for LP}},
\end{equation}
where $\kappa\equiv 6^{2n^*}\xi_i/(3n^*\Gamma(2n^*))$ is a coefficient only related to the atom/ion species and $\hat{w}\equiv w/w_0$. $\epsilon_\psi\equiv 3E_p/2\xi_0$ is a characteristic parameter which, for typical laser intensity near the ionization threshold, meets $\epsilon_\psi\ll 1$. For example, in a standard laser-driven injection scenario with the laser wavelength $\lambda_0=0.8\ \mu{\rm m}$ and normalized vector potential $a_0=2$ considering ${\rm N}^{5+}$ as the injection source, we have $\epsilon_\psi\sim0.09$. In the case that helium serves as injection source, a 800 nm laser with $a_0\sim0.03$ yields $\epsilon_\psi\sim0.1$. Therefore, the derivation in this section is on the assumption $\epsilon_\psi\ll1$ by default. It is notable that the exponential factor $\exp(-\hat{w}/\epsilon_\psi)$ in Eq. (\ref{eq:onaxis_iondeg}) makes a significant diminution in ionization degree as $\hat{w}$ changes not too much, which makes it possible to terminate the ionization and injection through laser spot evolution.

\begin{figure}
    \centering
    \includegraphics[width=10cm]{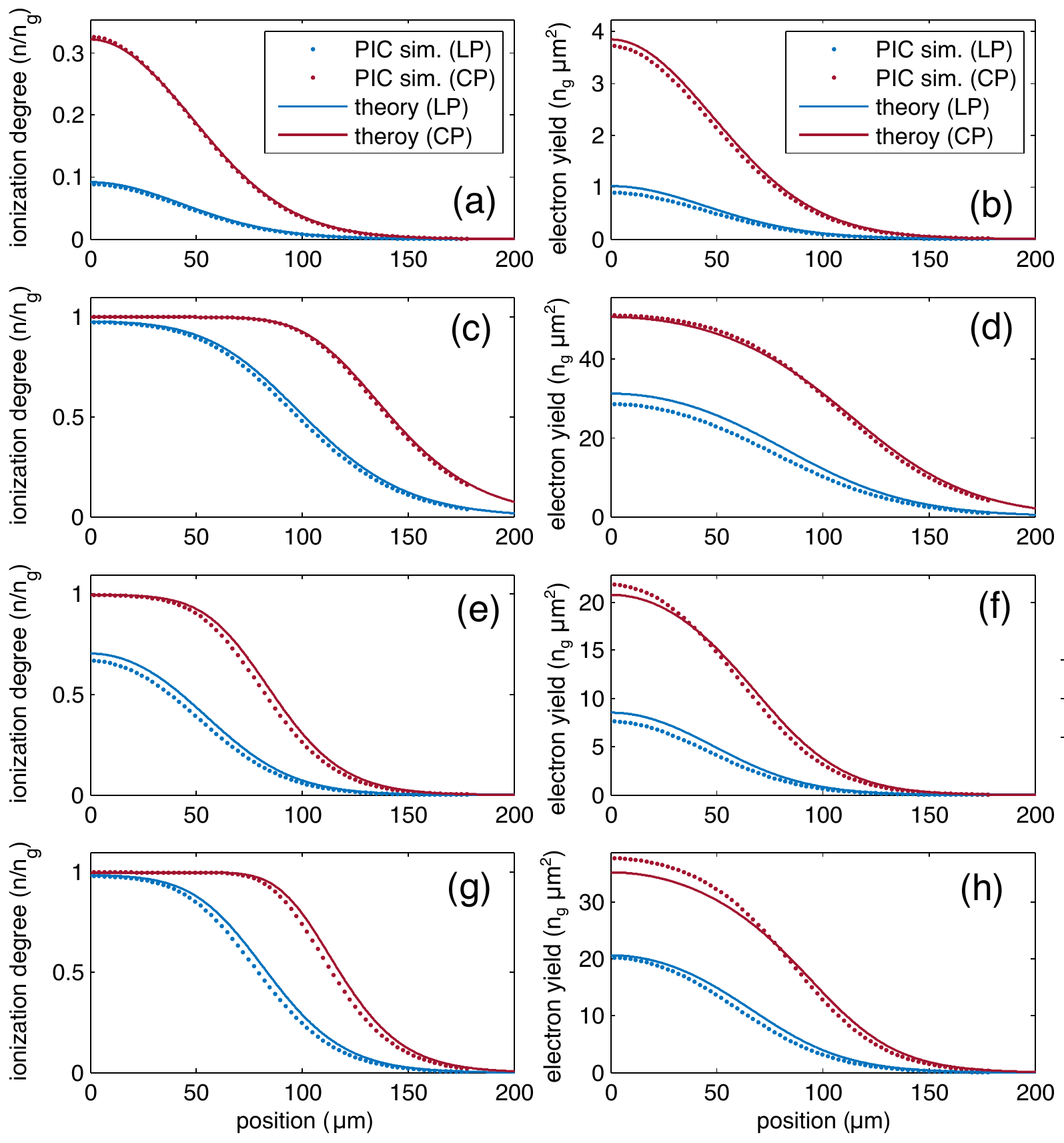}
    \caption{\label{fig:ion_deg}Ionization degree (left column) and electron yield (right column) along the direction the laser propagates. The laser pulse with $\lambda_0=800$ nm is focused at $z=0\ \mu$m. (a-b) $a_0=0.02$ for He, (c-d) $a_0=0.03$ for He, (e-f) $a_0=1.8$ for N$^{5+}$, (g-h) $a_0=2.0$ for N$^{5+}$.}
\end{figure}

Besides the ionization degree, electron yield, \emph{i.e.} the electrons released by laser per unit distance along the laser propagating direction, is also our concern. Integrating $n/n_g$ (see \ref{appendix:B}) over the transverse coordinates gives the electron yield
\begin{equation}\label{eq:electron_yield}
    \frac{dN}{dz}=\pi n_g \epsilon_\psi \hat{w}w_0^2[\gamma_e-{\rm Ei}(-I_0)+\ln I_0],
\end{equation}
where $\gamma_e$ is the Euler-Mascheroni constant and ${\rm Ei}(x)$ is the exponential integral. $I_0$ is the on-axis ionization degree given in Eq. (\ref{eq:onaxis_iondeg}). In weak ionization limit $I_0\ll1$ and full ionization limit $I_0\gg1$, the electron yield reduces to
\begin{equation}\label{eq:electron_yield_reduce}
    \frac{dN}{dz}=
    \cases{\pi n_g \epsilon_\psi \hat{w}w_0^2 I_0, \quad \mbox{for weak ionization}, \\
        \pi n_g \epsilon_\psi \hat{w}w_0^2 \ln I_0, \quad \mbox{for full ionization}.}
\end{equation}
In the proposed injection scheme, the target atoms/ions are usually far from fully-ionized hence we only consider the weak ionization limit. Since the injection distance has not be explicitly defined in existing literatures, we firstly define $D_{\rm inj}$ as the distance where $d_z N$ falls to $e^{-1}$ that at the focal plane. Performing a standard order-by-order expansion to $d_z N(D_{\rm inj})=e^{-1}d_z N(0)$ in terms of $\epsilon_\psi$ yields the laser spot size at $D_{\rm inj}$
\begin{equation}\label{eq:spotsize_at_dinj}
    \hat{w}(D_{\rm inj})=1+\epsilon_\psi+c(n^*)\epsilon_\psi^2,
\end{equation}
where $c(n^*)=2n^*-1$ for LP laser and $2n^*-\frac{1}{2}$ for CP laser. As the concrete expression of $D_{\rm inj}$ depends on how the laser spot evolves during the ionization process, to explicitly solving $D_{\rm inj}$ it is rational to assume the evolution of $\hat{w}$ around the focal plane is analogous to that at free space, \emph{i.e.} $\hat{w}=\sqrt{1+z^2/z_R^2}$, because the mismatched plasma channel hardly help guide the laser when it's tightly focused and the self-focusing condition $P/P_c>1$ is far from being reached. Under this assumption, we have (see \ref{appendix:C})
\begin{equation}\label{eq:inj_distance}
    D_{\rm inj}\approx \sqrt{2\epsilon_\psi}z_R.
\end{equation}
As a matter of fact, if the evolution of $\hat{w}$ is known, $D_{\rm inj}$ can be derived according to Eq. (\ref{eq:spotsize_at_dinj}). In a mismatched plasma channel, the oscillation of $\hat{w}$ has a general solution and we can perform an expansion around where the laser get focused, obtain the effective Rayleigh length $\tilde{z}_R$ (see \ref{appendix:C}). Replacing $z_R$ in Eq. (\ref{eq:inj_distance}) with $\tilde{z}_R$ gives the injection distance for plasma channel scenario.

\begin{figure}
    \centering
    \includegraphics[width=7cm]{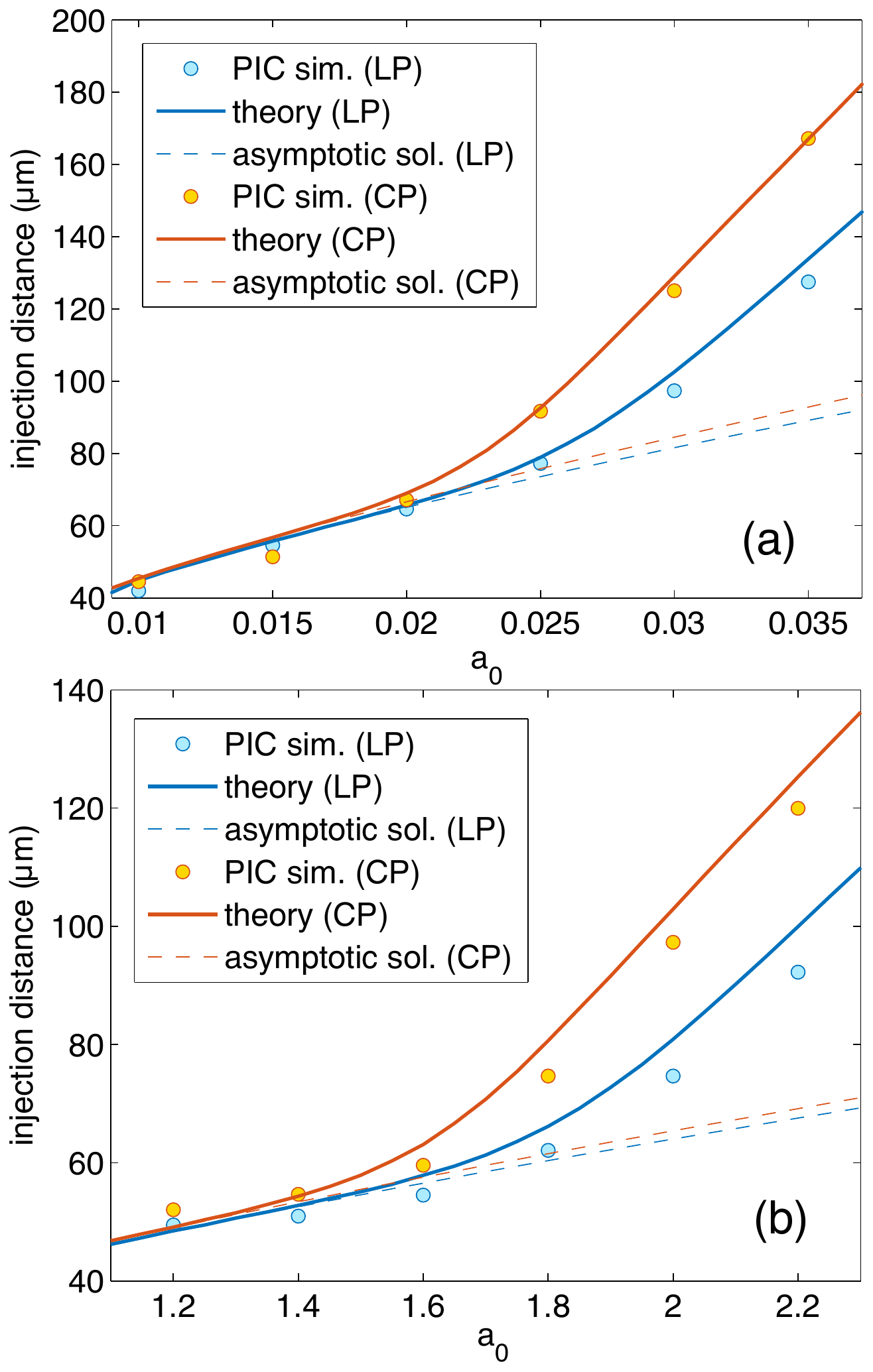}
    \caption{\label{fig:inj_dist}Injection distance under different $a_0$ for (a) He and (b) N$^{5+}$. The asymptotic solution refers to Eq. (\ref{eq:inj_distance}). Orange and blue lines denote CP and LP cases, respectively. In both He and N$^{5+}$ cases, the parameters of laser pulses are set to be fixed value, $\lambda_0=0.8\ \mu$m, $w_0=6\ \mu$m and $\tau=30$ fs.}
\end{figure}

Fig. \ref{fig:ion_deg} shows the comparison between the particle-in-cell (PIC) simulation and theory. Eq. (\ref{eq:onaxis_iondeg}) and (\ref{eq:electron_yield}) are verified for He and N$^{5+}$ and different laser normalized vector potential $a_0$ with fixed $w_0=6\ \mu$m and fixed pulse duration $\tau=30$ fs. Red and blue lines correspond to CP laser and LP laser. As seen in Fig. \ref{fig:ion_deg}, for different tested atom/ion species under different $a_0$, our theory gives an excellent description. Fig. \ref{fig:inj_dist} shows the comparison between theory and PIC simulation for injection distance under different $a_0$. Likewise, the theory gives a good prediction of injection distance in He and N$^{5+}$ cases.

In addition, integrating $d_zN$ along $z$ gives the number of ionized electrons
\begin{equation}\label{eq:ion_electrons}
    N=\sqrt{2}\frac{\kappa}{c}\pi^{\frac{3}{2}}n_g\sigma_\xi w_0^2 z_R \exp\left(-\frac{1}{\epsilon_\psi}\right) \times
    \cases{\sqrt{\pi}\epsilon_\psi^{-2n^*},\ \mbox{for CP}, \\
        \sqrt{2}\epsilon_\psi^{\frac{1}{2}-2n^*},\ \mbox{for LP}.}
\end{equation}
This expression gives the total number of electrons ionized along the laser's path. In some special injection schemes in which almost all the electrons are captured by wakefield, Eq. (\ref{eq:ion_electrons}) gives a good estimation for the captured beam charge.

Noting $\epsilon_\psi\propto E_p$, Eq. (\ref{eq:inj_distance}) tells that $D_{\rm inj}$ is insensitive to laser intensity but sensitive to focal waist thus $D_{\rm inj}$ can be adjusted mainly by changing $w_0$ and certain level of the laser intensity fluctuation is tolerated. Due to $N\propto w_0^2 z_R\propto w_0^4$, ionized charge may varies significantly as $w_0$ changes. The factor $\exp(-1/\epsilon_\psi)\epsilon_\psi^{-2n^*}$ at $\epsilon_\psi\ll1$ ensures the ionized charge fluctuation is acceptable to some extent. For example, 5\% laser intensity fluctuation around $a_0=2$ ($\epsilon_\psi\sim0.1$ for $\lambda_0=0.8\ \mu{\rm m}$) when ionizing N$^{5+}$ only leads to 1.3\% fluctuation in $D_{\rm inj}$ (or $\Delta\gamma_0$) and 29\% in ionized charge.

\section{PIC simulation of near-ionization-threshold injection}

To examine the mechanism and feasibility of NIT injection, we performed 3D PIC simulations using code \textsc{osiris} \cite{fonseca2002}. Hereon, we just present a typical example from a series of simulations. Throughout this section, we employ a pre-plasma channel with parabolic transverse density profile which serves to guide laser pulses. The pre-plasma channel can be readily fabricated by some laser-ionization techniques like ignitor-heater scheme \cite{volfbeyn1999,xiao2004} or single sub-picosecond laser pulse scheme \cite{lemos2013}. The pre-plasma channel is doped with nitrogen, a frequently-used candidate in ionization injection since it has a wide ionization potential gap between K-shell and L-shell electrons (97.9 eV for N$^{4+}$ and 552 eV for N$^{5+}$). The outer five electrons of nitrogen are stripped by the prepulse, forming the plasma wakefield, while the 6th electron is freed by the main pulse, serving as the injection source.

The simulations were performed in Cartesian coordinate defining the laser pulse propagates in \textit{z}-direction. The dimension of simulation window is $76.3\times76.3\times50.9\ \mu{\rm m}^3$ containing $400\times400\times2400$ cells in total. On the consideration of shortening the injection distance as much as possible, the laser pulse should have small $w_0$ and $a_0$, fulfilling a short $z_R$ and meeting the weak ionization condition according to Eq. (\ref{eq:electron_yield}). On the other hand, the requirement of injecting sufficient charge and exciting a wakefield strong enough to capture and accelerate electrons set a lower limit for selecting $w_0$ and $a_0$. We made a tradeoff between the above considerations, setting $a_0=2,\ w_0=6\ \mu$m and $\tau=30$ fs. The on-axis plasma density $n_p=1\times10^{18}$ cm$^{-3}$, which is purposely set such low to weaken self-focusing effect. The radius $r_0$ and depth $\Delta n$ of plasma channel are set to be 7.6 $\mu$m and $1\times10^{18}$ cm$^{-3}$. Under this condition, one can verify that the matched plasma channel condition $\Delta n w_0^4=\Delta n_c r_0^4$, under which the laser propagates with a constant spot size, is violated, where $n_c=(\pi r_e r_0^2)^{-1}$ and $r_e$ the classic electron radius. In such mismatched plasma channel, laser disperses rapidly, hereby shortening the injection distance. The concentration of N$^{5+}$ is 50\%, \emph{i.e.} $5\times10^{17}$ cm$^{-3}$, which is much higher than conventional ionization injection schemes. The purpose of doing this is mainly to compensate reduction of injected charge resulting from injection distance shortening.

\begin{figure}
    \centering
    \includegraphics[width=16cm]{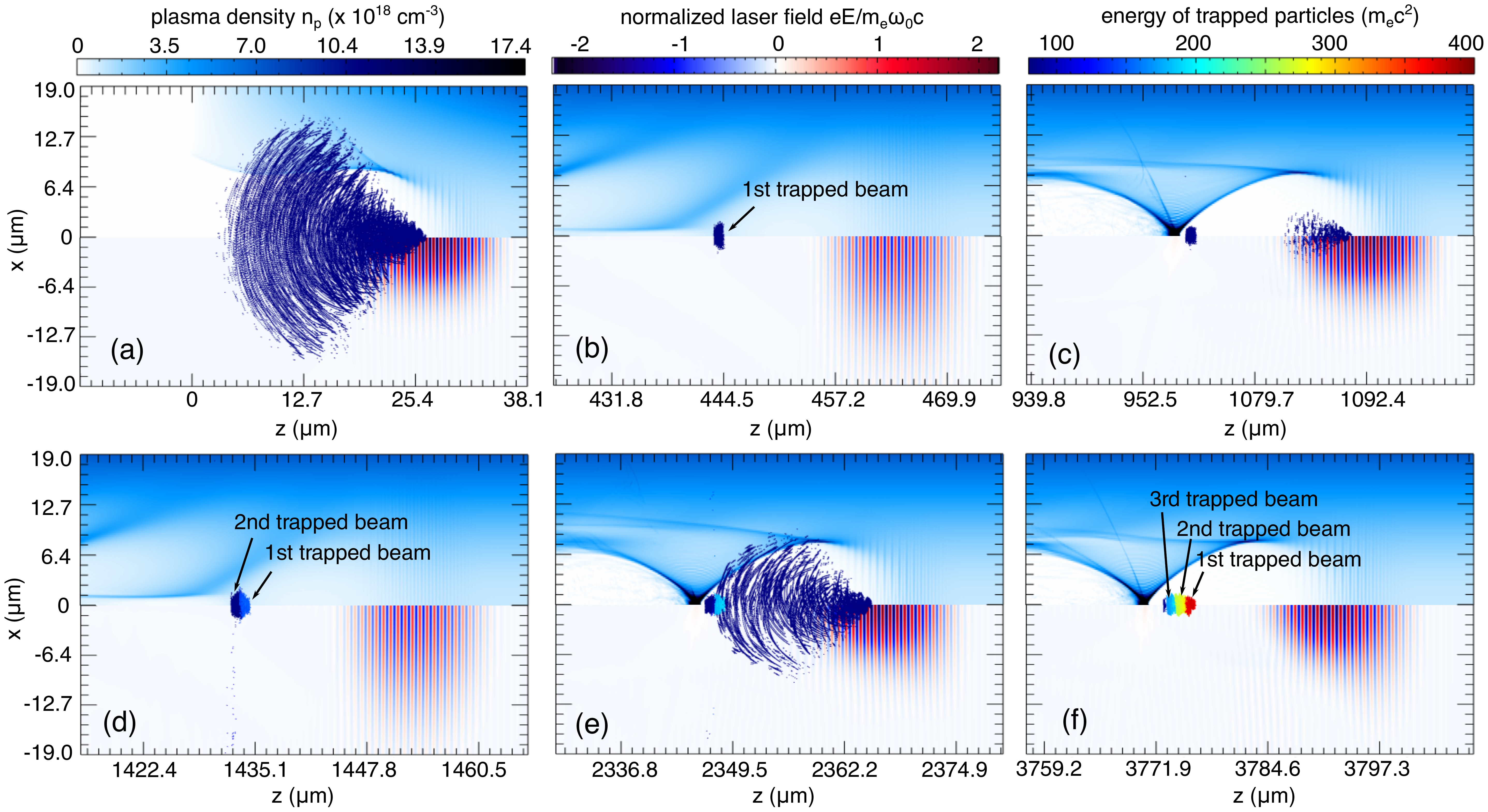}
    \caption{\label{fig:inj_process}Injection and acceleration process of NIT injection scheme. Each subfigure consists of an upper half showing the plasma density distribution (colored blue) and a lower half showing the laser field (colored blue-red). (a) The moment laser just enters into the plasma channel, releasing a large fraction of N$^{5+}$ electrons (dark blue). (b) The moment laser disperse to its largest spot size. The injection has ceased and a beam has captured by wakefield. (c) The laser reaches its focal intensity and starts to ionize electrons again. (d) Laser disperse again and the second beam is trapped behind the first beam. (e) The laser gets focused and injection occurs for the third time. (f) Multi-bunch consisting of three sub-beams (colored red, yellow and light blue) with different energy forms ultimately.}
\end{figure}

Fig. \ref{fig:inj_process} illustrates the whole process of injection and acceleration. In Fig. \ref{fig:inj_process}(a), a laser pulse focused at $z=38\ \mu$m enters into a pre-plasma channel, excite a highly nonlinear wakefield and start to deliver electrons of N$^{5+}$ into the wakefield. As the mismatched channel exerts little guide effect on the tightly focused laser at this moment, the laser pulse disperse rapidly with its peak intensity falls to nearly a quarter of the focal intensity, halting the ionization. A small fraction of ionized electrons are trapped in the wakefield as depicted in Fig. \ref{fig:inj_process}(b). When the laser spot expands to a large size, the confinement and guide effect exerted by the plasma channel becomes significant. The laser stops dispersing and turn to focusing. Once reaching its focal intensity, the laser switches on a new round of ionization and injection, as shown in Fig. \ref{fig:inj_process}(c-d). Because of the phase slippage, the second injected beam is located behind the first one. With the oscillation of laser spot in the plasma channel, the third and fourth round take place likewise, injecting the third beam (the fourth beam has only a negligible fraction of particles, which can hardly observed in the figure.) as shown in Fig. \ref{fig:inj_process}(e-f). Ultimately, after acceleration within 3.8 mm plasma, we obtain three mono-energetic electron beams with separated energy.

\begin{figure*}
    \centering
    \includegraphics[width=16cm]{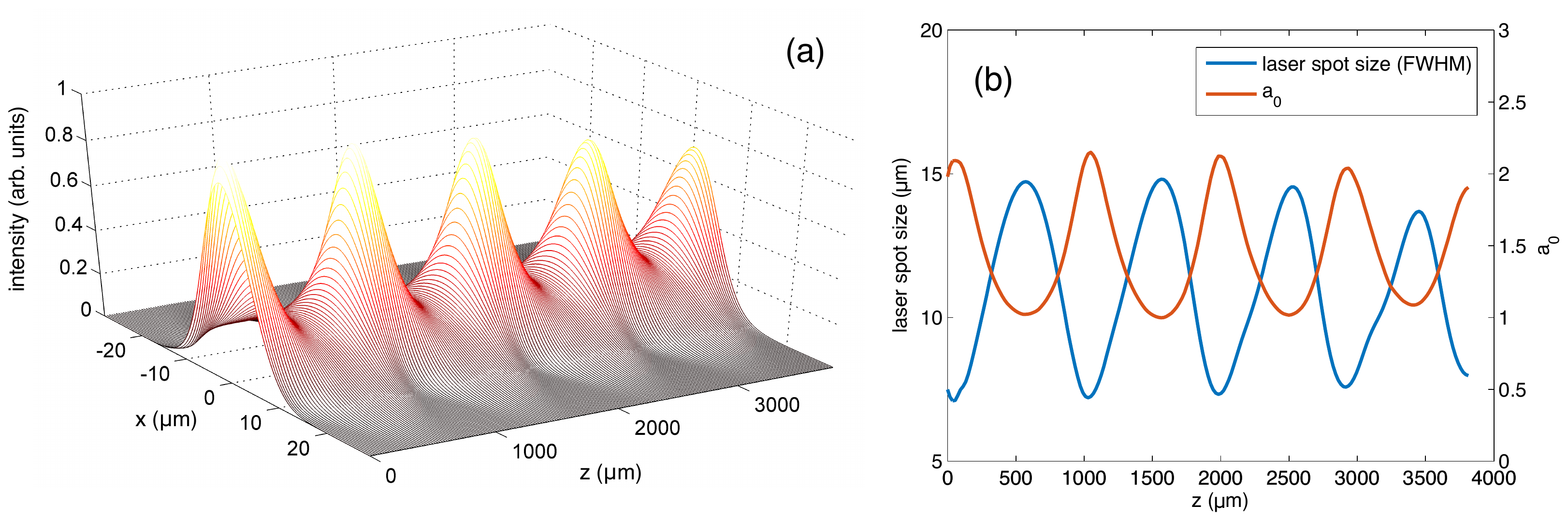}
    \caption{\label{fig:laser_evolution}Evolution of laser pulse in the mismatched plasma channel. (a) Evolution of (normalized) transverse intensity distribution. (b) Oscillation of laser's spot size (FWHM) (blue line) and $a_0$ (red line) as the laser propagates.}
\end{figure*}

Fig. \ref{fig:laser_evolution} shows the evolution of laser spot and intensity. From Fig. \ref{fig:laser_evolution}(a), one can clearly see that the laser's transverse intensity envelope always maintains a gaussian-like distribution, only with peak intensity and spot size oscillating. According to laser's evolution, one can locate the injection position at around $z=0\ \mu$m (38 $\mu$m exactly), \textit{z}=1 mm and \textit{z}=2 mm, as indicated by Fig. \ref{fig:laser_evolution}(b). Because of the depletion during the propagation, the focal intensity of the laser pulse cannot retrieve its initial value and decrease every time it gets focused. Hence, during its fourth focusing, the laser is depleted and no longer can releases enough electrons to generate the fourth beam, as we mentioned above.

\begin{figure}
    \centering
    \includegraphics[width=8cm]{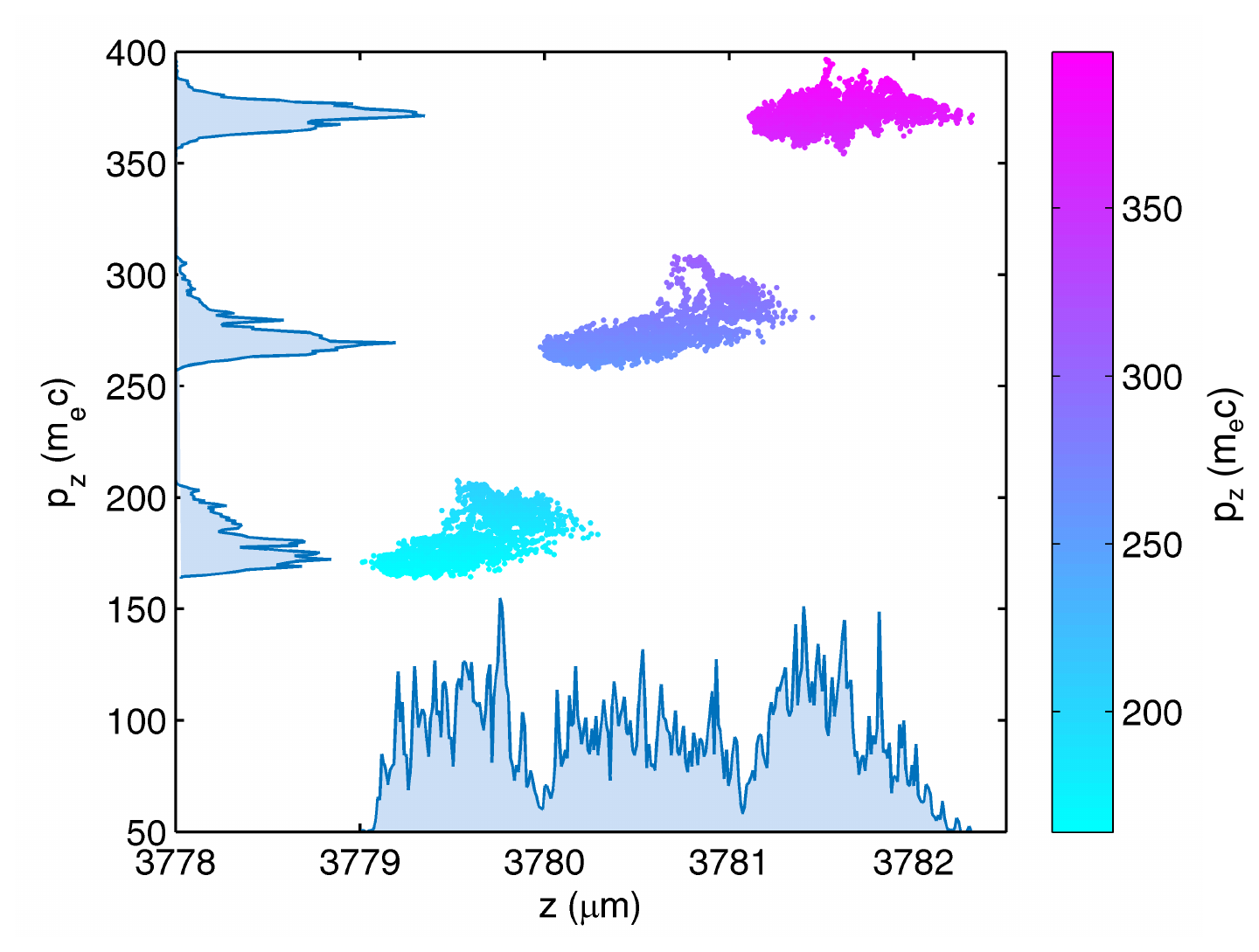}
    \caption{\label{fig:x1p1}Longitudinal phase-space distribution of the trapped multi-bunch.}
\end{figure}

Because of the uniqueness of the proposed injection scheme, the trapped beams have some unique features accordingly. The output beam parameters are listed in Table \ref{tab:beam_para}. Firstly, since the laser spot oscillates at a nearly fixed frequency in the plasma channel, the intervals of injection position of each beam are almost identical, which leads to beams are nearly evenly spaced along \textit{z} and $p_z$ axis in the phase-space, as presented in Fig. \ref{fig:x1p1}. Secondly, just like the core idea in this paper that obtaining low energy spread by reducing injection distance, the (first) injected beam has a narrow energy spread (rms) of 2.7 MeV corresponding to a relative energy spread down to 1.4\%. The injection distance is estimated to be 60 $\mu$m which agrees well with simulation results.

\begin{table}
\caption{\label{tab:beam_para}Beam parameters of the presented results.}
\begin{indented}
\item[]\begin{tabular}{llll}
    \br
     & 1st beam & 2nd beam & 3rd beam \\
    \mr
    peak energy (MeV) & 190 & 140 & 92 \\
    energy spread (MeV) & 2.7(1.4\%) & 4.5(3.2\%) & 4.8(5.2\%) \\
    charge (pC) & 1.7 & 1.6 & 1.5 \\
    emittance$^{\rm a}$ ($\mu$m) & $1.17\times0.37$ & $0.76\times0.23$ & $0.51\times0.22$ \\
    rms beam length (fs) & 1.2 & 1.3 & 1.7 \\
    peak current (kA) & 0.73 & 0.52 & 0.68 \\
    \br
\end{tabular}
\item[] $^{\rm a}$ The normalized emittance is quantified using $\epsilon_n=\sqrt{\langle x^2\rangle \langle p_x^2\rangle-\langle xp_x\rangle^2}/m_ec$.
\end{indented}
\end{table}

\begin{figure}
    \centering
    \includegraphics[width=16cm]{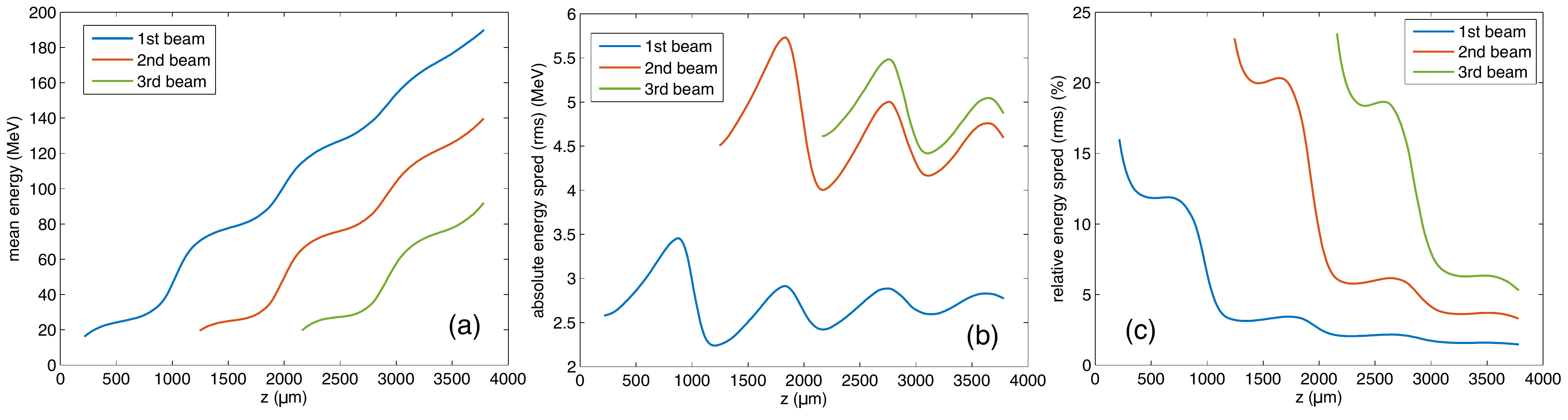}
    \caption{\label{fig:beam_evol}Variation of (a) mean energy, (b) absolute rms energy spread and (c) relative rms energy spread during the acceleration.}
\end{figure}

We traced how the energy and energy spread of trapped beams change during the acceleration, as shown in Fig. \ref{fig:beam_evol}. Other than conventional acceleration process, the energy gain curves present a rising trend in stages. This is because the evolution of laser leads to the wakefield evolves acutely, converting between highly nonlinear regime and weakly nonlinear regime. The variation of acceleration gradient results in unique energy gain curves in Fig. \ref{fig:beam_evol}(a) as well as the oscillation of absolute energy spread in Fig. \ref{fig:beam_evol}(b). For long-distance acceleration, extra energy spread (energy chirp) may be introduced since the nonuniformity of acceleration field felt by the beams. In our simulations, the beams' length (rms) is no more than 2 fs only occupying an extremely narrow accelerating phase, which may tremendously limit the growth of absolute energy spread. As a consequence of that, the relative energy spread (see Fig. \ref{fig:beam_evol}(c)) continuously decreases during the acceleration and further acceleration makes energy spread less than 1\% quite promising.

\section{Conclusion}

In this paper, we have proposed an enhanced ionization injection scheme using a tightly focused laser pulse with intensity near the ionization potential to trigger the injection process in a mismatched pre-plasma channel. The whole injection and acceleration process were examined through multi-dimensional PIC simulation using \textsc{osiris}. The core idea of the proposed scheme is to lower the energy spread of trapped beams by shortening the injection distance. Firstly, we have derived the expression of ionization degree, electron yield, injection distance and ionized charge in the case of bi-gaussian distribution laser pulse. Through this theory, we can give a precise estimation of injection distance and optimize the initial energy spread by tuning laser parameters ($w_0,\ \tau$ and $a_0$). This theoretical model is not only restricted to the proposed injection scheme, but also can be extended to any ionization process triggered by single laser pulse. Secondly, we carried out PIC simulations to explore the injection and acceleration process, and found out some unique features of this scheme. The feasibility and adjustability have been verified. Spatial and spectral spacing of the injected multi-beam can be changed through fine-tuning laser and plasma channel parameters. By optimizing the simulation parameters, beams with energy spread (rms) down to 1.4\% with charge 1.7 pC and normalized emittance down to $1.17\times0.37\ \mu$m can be extracted. Further acceleration may continue bringing down the relative energy spread. Fine-tuning parameters may make energy spread less than 1\% quite promising, which may have potential in applications related to future coherent light source driven by laser-plasma accelerators.

\section*{Acknowledgment}

This work is supported by National Natural Science Foundation of China Grant No.11375006, No. 11425521, No. 11535006, No. 11475101 and No.11175102. Simulations are performed on Hoffman cluster at UCLA and Hopper cluster at NERSC.

\appendix

\section{\label{appendix:A}Derivation of ionization degree}

In this appendix we derive the formula of ionization degree. Before performing the derivation we firstly present a mathematical skill that will be repeatedly applied in the derivation below. This skill called Laplace asymptotic expansion (LAE) is aimed to approximately calculate integrals of the form
\begin{equation}\label{eq:laplace_expansion1}
    I(x)=\int_{-\infty}^{+\infty}f(t)e^{x\phi(t)}dt, \quad {\rm for}\ x\gg0.
\end{equation}
As the integral kernel $e^{x\phi(t)}$ diminishes rapidly near the maxima of $\phi(t)$, the major contribution to the integral comes from the neighborhood of $t=c$. Set $f(t)\simeq f(c)$ and $\phi(t)\simeq\phi(c)+\frac{1}{2}\phi''(c)(t-c)^2$, we have
\begin{equation}\label{eq:laplace_expansion2}
    I(x)\sim f(c)e^{x\phi(c)}\sqrt{\frac{2\pi}{|\phi''(c)|x}}.
\end{equation}
See reference \cite{bender1999advanced} for more detailed and rigorous proof.

\subsection{Circularly Polarized Laser}

For a CP laser with bi-gaussian envelope, the electric field is
\begin{equation}\label{eq:efield_cp}
    E=E_p\frac{w_0}{w}\exp\left(-\frac{r^2}{w^2}-\frac{\xi^2}{\sigma_\xi}\right).
\end{equation}
Insert Eq. (\ref{eq:efield_cp}) into Eq. (\ref{eq:adkmodel}) and integrate over $t$, we have
\begin{equation}
    \eqalign{I(r,z) &= \int_{-\infty}^{+\infty}W_{\rm ADK}(t)dt \cr
    &= \frac{\kappa}{c}\int_{-\infty}^{+\infty}\left(\frac{\hat{w}}{\epsilon_\psi f_r(r)f_\xi(\xi)}\right)^{2n^*-1} \exp\left[-\frac{\hat{w}}{\epsilon_\psi f_r(r)f_\xi(\xi)}\right]d\xi,}
\end{equation}
where $f_r(r)\equiv\exp(-r^2/w^2)$ and $f_\xi(\xi)\equiv\exp(-\xi^2/\sigma_\xi^2)$. Because of $\epsilon_\psi\ll 1$ it is valid to perform Laplace asymptotic expansion around where the peak field is reached, \emph{i.e.} $\xi=0$, which yields
\begin{equation}\label{eq:iondegree_cp}
    I(r,z)\simeq \frac{\kappa}{c}\sqrt{\pi\sigma_\xi^2} \left[\frac{\hat{w}}{\epsilon_\psi f_r(r)}\right]^{2n^*-\frac{3}{2}} \exp\left[-\frac{\hat{w}}{\epsilon_\psi f_r(r)}\right].
\end{equation}
Setting $r=0$ gives the on-axis ionization degree as shown in Eq. (\ref{eq:onaxis_iondeg}).

\subsection{Linearly Polarized Laser}

Compared with CP laser field, it would be more complicated and tedious to deal with LP laser mainly because of the appearance of oscillation term.
\begin{equation}\label{eq:efield_lp}
    E=E_p\frac{w_0}{w}\exp\left(-\frac{r^2}{w^2}-\frac{\xi^2}{\sigma_\xi}\right)\cos k\xi.
\end{equation}
Supposing $\sigma_\xi\gg k^{-1}$, a reasonable strategy is to integrate $W_{\rm ADK}$ over a single electric field period firstly and sum up the contribution of each period to the integral. For $j^{\rm th}$ period, integrate over the range $[(j-\frac{1}{2})\frac{\pi}{k},(j+\frac{1}{2})\frac{\pi}{k}]$

\begin{equation}
    \fl I_j=\frac{\kappa}{c}\int_{(j-\frac{1}{2})\frac{\pi}{k}}^{(j+\frac{1}{2})\frac{\pi}{k}} \left[\frac{\hat{w}}{\epsilon_\psi f_r(r)f_\xi(\xi)|\cos k\xi|}\right]^{2n^*-1} \exp\left[-\frac{\hat{w}}{\epsilon_\psi f_r(r)f_\xi(\xi)|\cos k\xi|}\right]d\xi,\quad j=0,\pm1,\pm2...
\end{equation}
Expand $|\cos k\xi|^{-1}$ around $\xi=j\pi/k$ and apply LAE method again, we get $j^{\rm th}$ integral
\begin{equation}
    I_j\simeq \frac{\kappa}{c}\sqrt{2\pi}k^{-1} \left[\frac{\hat{w}}{\epsilon_\psi f_r(r)f_\xi^j}\right]^{2n^*-\frac{3}{2}} \exp\left[-\frac{\hat{w}}{\epsilon_\psi f_r(r)f_\xi^j}\right], \quad j=0,\pm1,\pm2...
\end{equation}
where $f_\xi^j\equiv f_\xi(\frac{j\pi}{k})$. Summation on $j$ gives the complete integral

\begin{equation}\label{eq:iondegree_lp}
    \eqalign{I&=\sum_j I_j=\frac{\kappa}{c}\sqrt{2\pi}k^{-1} \sum_j \left[\frac{\hat{w}}{\epsilon_\psi f_r(r)f_\xi^j}\right]^{2n^*-\frac{3}{2}} \exp\left[-\frac{\hat{w}}{\epsilon_\psi f_r(r)f_\xi^j}\right] \\
    &\simeq \frac{\kappa}{c}\sqrt{\frac{2}{\pi}} \int_{-\infty}^{+\infty} \left[\frac{\hat{w}}{\epsilon_\psi f_r(r)f_\xi(\xi)}\right]^{2n^*-\frac{3}{2}} \exp\left[-\frac{\hat{w}}{\epsilon_\psi f_r(r)f_\xi(\xi)}\right]d\xi \\
    &\simeq \frac{\kappa}{c}\sqrt{2}\sigma_\xi \left[\frac{\hat{w}}{\epsilon_\psi f_r(r)}\right]^{2n^*-2} \exp\left[-\frac{\hat{w}}{\epsilon_\psi f_r(r)}\right]}
\end{equation}
Here, we have applied Laplace asymptotic expansion again to the last approximately equal sign. Setting $r=0$ gives the on-axis ionization degree as shown in Eq. (\ref{eq:onaxis_iondeg}).

Having obtained the ionization degree under weak ionization limit, the general form of ionization degree $n/n_g$ can be written as in Eq. (\ref{eq:ion_degree}).

\section{\label{appendix:B}Derivation of electron yield}

In this appendix we derive the electron yield, \emph{i.e.} the number of ionized electron per unit distance along the laser propagation direction. The electron yield can be attained through integrating the ionization degree $n/n_g$ over transverse coordinates $x,\ y$
\begin{equation}
    \frac{dN}{dz}=\int\int 1-\exp(-I(r,z))dxdy.
\end{equation}
To get the explicit form of the integral, we apply series expansion to the integrand $1-\exp(-I)=\sum_{n=1}^{+\infty}(-1)^{n+1}I^n/n!$ and integrate each term of the infinite series
\begin{equation}
    \eqalign{\fl s_n=\int\int I^n dxdy = \left(\frac{\kappa}{c}\sqrt{\pi\sigma_\xi^2}\right)^n \int\int \left[\frac{\hat{w}}{\epsilon_\psi f_r(r)}\right]^{(2n^*-\frac{3}{2})n} \exp\left[-\frac{n\hat{w}}{\epsilon_\psi f_r(r)}\right] dxdy \cr
    \simeq \frac{\pi\epsilon_\psi\hat{w}w_0^2}{n} \left[\frac{\kappa}{c}\sqrt{\pi\sigma_\xi^2} \left(\frac{\hat{w}}{\epsilon_\psi}\right)^{2n^*-\frac{3}{2}} \exp\left(-\frac{\hat{w}}{\epsilon_\psi}\right)\right]^n = \frac{\pi\epsilon_\psi\hat{w}w_0^2}{n}I_0^n.}
\end{equation}
Here, although we have taken the case of CP laser for example to perform the integrating, the form of $s_n$ is completely same for LP laser. Substituting $s_n$ into the formula $d_zN=\sum_{n=1}^\infty(-1)^{n+1}s_n/n!$ leads to a differential equation
\begin{equation}
    \frac{\partial}{\partial I_0}\left(\frac{dN}{dz}\right)=\frac{\pi\epsilon_\psi\hat{w}w_0^2}{I_0}(1-e^{-I_0}), \quad \left.\frac{dN}{dz}\right|_{I_0=0}=0.
\end{equation}
The solution can be written as
\begin{equation}
    \frac{dN}{dz}=\pi\epsilon_\psi\hat{w}w_0^2[\gamma_e-{\rm Ei}(-I_0)+\ln I_0]
\end{equation}
where $\gamma_e$ the Euler-Mascheroni constant and {\rm Ei}(x) the exponential integral. In the weak ionization limit $I_0\ll1$, applying the asymptotic behavior ${\rm Ei}(x)\rightarrow\gamma_e+\ln|x|+x$ when $x\rightarrow0$ we have
\begin{equation}
    \frac{dN}{dz}=\pi\epsilon_\psi\hat{w}w_0^2 I_0, \quad {\rm for}\ I_0\ll1.
\end{equation}
In full ionization limit, \emph{i.e.} $I_0\gg1$, ${\rm Ei}(-I_0)$ vanishes rapidly and $\gamma_e$ can be ignored. Thus we have
\begin{equation}
    \frac{dN}{dz}=\pi\epsilon_\psi\hat{w}w_0^2 \ln I_0, \quad {\rm for}\ I_0\gg1.
\end{equation}

\section{\label{appendix:C}Derivation of injection distance and ionized charge}

In this appendix, we derive the injection distance under weak ionization limit and the total ionized charge number during injection. Defining the injection distance $D_{\rm inj}$ to be position where the electron yield falls to $1/e$ that at the focal plane, we need to solve the equation
\begin{equation}
    \hat{w}^{c(n^*)}e^{-\hat{w}/\epsilon_\psi}=e^{-1/\epsilon_\psi-1},
\end{equation}
where $c(n^*)=2n^*-1$ for LP laser and $c(n^*)=2n^*-\frac{1}{2}$ for CP case. Due to the appearance of the characteristic small parameter $\epsilon_\psi$, letting $u\equiv(\hat{w}-1)/\epsilon_\psi$ we seek for the perturbative solution $u=u^{(0)}+\epsilon_\psi u^{(1)}+...$. Solving the equation order by order gives the laser spot size at $D_{\rm inj}$
\begin{equation}
    \hat{w}(D_{\rm inj})\simeq 1+\epsilon_\psi+c(n^*)\epsilon_\psi^2.
\end{equation}
Extracting the form of $D_{\rm inj}$ depends on how the laser spot evolves. In an unmatched plasma channel provided the self-focusing effect is neglectable, the laser evolves like in free space, \emph{i.e.} $\hat{w}=\sqrt{1+z^2/z_R^2}$. Therefore, only keeping the leading order of $\epsilon_\psi$, $\hat{w}(D_{\rm inj})=\sqrt{1+D_{\rm inj}^2/z_R^2}$ gives
\begin{equation}\label{eq:inj_distance_app}
    D_{\rm inj}=\sqrt{2\epsilon_\psi}z_R.
\end{equation}
Integrating $d_zN$ along $z$-axis yields the total ionized electron number
\begin{equation}\label{eq:ion_charge_app}
    N=\pi\epsilon_\psi w_0^2\int_{-\infty}^{+\infty}\hat{w}(z)I_0(z)dz.
\end{equation}
Insert $\hat{w}=\sqrt{1+z^2/z_R^2}$ and Eq. (\ref{eq:onaxis_iondeg}) into Eq. (\ref{eq:ion_charge_app}) and apply LAE around $z=0$, we derive
\begin{equation}
N=\sqrt{2}\frac{\kappa}{c}\pi^{\frac{3}{2}}n_g\sigma_\xi w_0^2 z_R \exp\left(-\frac{1}{\epsilon_\psi}\right) \times
\cases{\sqrt{\pi}\epsilon_\psi^{-2n^*},\ \mbox{for CP}, \\
        \sqrt{2}\epsilon_\psi^{\frac{1}{2}-2n^*},\ \mbox{for LP}. \\}
\end{equation}

Above derivation is based on the assumption of laser drifting in free space, which, in an mismatched plasma channel, is a very good approximation. In fact, to be more convincing, we can consider the evolution of laser spot size in a parabolic density channel of the form $n=n_0+\Delta nr^2/r_0^2$, where $\Delta n$ and $r_0$ are channel depth and width respectively. The general solution \cite{esarey2009} is
\begin{equation}
    \hat{w}^2=\frac{1}{2}\left[1+\frac{\Delta n_c r_0^4}{\Delta n w_0^4} + \left(1-\frac{\Delta n_c r_0^4}{\Delta n w_0^4}\right)\cos(k_{\rm os}z) \right],
\end{equation}
where $\Delta n_c=(\pi r_er_0^2)^{-1}$ with $r_e$ being the classical electron radius is the critical channel depth and $k_{\rm os}=(2/z_R)(\Delta n/\Delta n_c)^{1/2}$ characterizes the oscillation period of spot size. For tightly focused lasers in unmatched channels $\Delta nw_0^4<\Delta n_c r_0^4$, considering the ionization only occurs near the focus, \emph{i.e.} $k_{\rm os}z\ll 1$, we expand the cosine term and get
\begin{equation}
    \hat{w}^2\simeq 1+2\frac{\Delta n}{\Delta n_c}\left(\frac{\Delta n_cr_0^4}{\Delta n w_0^4}-1\right)\frac{z^2}{z_R^2}
\end{equation}
It resembles the form of evolution in vacuum except for laser diffracting within an equivalent Rayleigh length $\tilde{z}_R=[2\left(r_0^4/w_0^4-\Delta n/\Delta n_c\right)]^{-1/2}z_R$. Substitute $\hat{z}_R$ for $z_R$ in Eq. (\ref{eq:inj_distance_app}) and (\ref{eq:ion_charge_app}), we will obtain the injection distance and ionized electron number of unmatched channel version.

\section*{References}

\bibliography{ref}

\end{document}